# Resonant Photon–Exciton Coupling in All-Semiconductor Heterostructures Composed of Silicon Nanosphere and Monolayer $WS_2$

*Hao Wang,[†,#] Jinxiu Wen,[†,#] Pu Liu,[§] Jiahao Yan,[§] Yu Zhang,[†,‡] Fei Liu,[†,‡] Juncong She,[†,‡] Jun Chen,[†,‡] Huanjun Chen,[†,‡,*] Shaozhi Deng,[†,‡,*] and Ningsheng Xu[†,‡,*]*

[†]State Key Laboratory of Optoelectronic Materials and Technologies, Guangdong Province Key Laboratory of Display Material and Technology, Sun Yat-sen University, Guangzhou 510275, China.

[‡]School of Electronics and Information Technology, Sun Yat-sen University, Guangzhou 510006, China.

[#]School of Physics, Sun Yat-sen University, Guangzhou 510275, China.

[§]School of Materials Science and Engineering, Guangzhou 510275, China.

*Address correspondence to stsxns@mail.sysu.edu.cn; stsdsz@mail.sysu.edu.cn; chenhj8@mail.sysu.edu.cn.

**ABSTRACT** Tailoring and enhancing the interaction between light and matter is of great importance for both fundamental researches and future photonic and optoelectronic applications. Due to their high exciton oscillator strength and large exciton binding energy, two-dimensional atomic semiconducting transition metal dichalcogenides have recently emerged as an excellent platform for the strong photon–exciton interaction by integrating with optically resonant cavities. Here, we propose an all-semiconductor system composed of individual silicon nanospheres and monolayer $WS_2$ and investigate the resonance coupling between these two constituents. By coating the silicon nanospheres with monolayer $WS_2$, we demonstrate the strong resonance coupling between the magnetic dipole mode and A-exciton, evidenced by an anticrossing behavior in the scattering energy diagram with a Rabi splitting of 77 meV. Compared with the plasmonic analogues, the resonance coupling in all-semiconductor heterostructure is much stronger and less sensitive to the spacing between the silicon nanosphere core and $WS_2$ shell. When the silicon nanospheres are placed onto the $WS_2$ monolayer with a point contact, resonance coupling manifested by the quenching dips in the scattering spectra can also be observed at ambient conditions, which involves only a few excitons. Finally, resonance coupling in the all-semiconductor heterostructure can be active controlled by temperature scanning. Our findings suggest that this all-semiconductor heterostructure can be exploited for future on-chip nanophotonics associated with strong light–matter interactions.

Over the last decade, two dimensional (2D) transition metal dichalcogenides (TMDs) have attracted much interest due to their significant potential for electronic and optoelectronic applications.[1, 2] In contrast to zero-bandgap graphene, several TMDs such as $MoS_2$, $MoSe_2$, $WS_2$ and $WSe_2$ are a class of semiconducting materials which have an inherent sizeable bandgaps in the near-IR spectral regime.[3] In particular, indirect-to-direct gap transition can occur at K (K′) points of the Brillouin zone when TMDs are thinned from bulk to the limit of monolayer,[4-6] resulting in enhanced photoluminesence (PL) and enabling access to valley degree of freedom.[5-7] The tightly bound excitons within the bandgap in TMD monolayers exhibit high oscillator strength and large banding energy and can be charged positively and negatively to form trions, which makes them promising for light–matter interaction at room temperature.[8-12] The interaction of 2D confined excitons with light can be further tailored from weak to strong coupling regime by integrating TMD monolayers with various optically resonant cavities.[13-27] For weak coupling, the absorption and PL emission of TMD monolayers can be strongly enhanced and modified by photonic crystal nanocavities[13] and plasmonic nanostructures.[14-18] Furthermore, Fano resonance as well as electromagnetically induced transparency (EIT) have been observed in the intermediate regime because of the coherent exciton–plasmon coupling.[18, 19] Once the coupling between the excitonic transitions and cavity photons is strong enough to overcome their decay and dissipation, the system reaches the strong coupling limit.[28-34] In this case, coherent and reversible energy exchange between the emitter and optical cavity leads to the formation of Rabi splitting and hybrid photon–exciton quasi-particles, namely, exciton-polaritons, which provides an approach for the study of quantum information processing, bosonic condensation effects and polariton lasers.[35-37] Very recently, strong photon–exciton coupling have been extensively reported in atomic semiconducting TMDs.[20-27] In addition, the strong interaction in TMDs can be actively controlled by electrostatic gating and thermal tuning,[23, 24, 27] which can hardly be achieved in quantum dots and dye molecules. More importantly, by taking advantage of the small mode volume of plasmonic nanostructures, strong coupling involving only a few excitons of 5 ~ 20 have been realized,[38] which however suffers from the large optical losses of plasmonic metals.

Recently, high-refractive-index dielectric nanoparticles made of semiconductor materials, such as silicon and germanium, have been proposed as low-loss building blocks for future nanophotonic devices.[39-41] Such dielectric nanoparticles exhibit strong optically induced electric and magnetic multipolar resonant modes in the visible and near-infrared spectral region, which in analogy to plasmonic metals, provide the opportunity to concentrate light into the subwavelength with strong electromagnetic near-filed enhancement and large far-field light scattering.[42-44] On the other hand, in comparison with their plasmonic counterparts, dielectric nanoparticles exhibit intrinsic low losses due to the small imaginary part of the refractive index and directional light scattering property resulting from the interference between the spectral overlapped electric (EDR) and magnetic (MDR) dipole resonance with comparable strengths.[45-48] The strong magnetic resonances can drastically enhance various optical processes, such as, nonlinear response,[49] Fano resonance,[50] photoemission,[51] and Raman spectroscopy.[52] In our recent work, we have demonstrated the resonance coupling between the magnetic response and excitonic transition manifested by quenching dips and splitting peaks on the scattering spectra in dielectric silicon nanospheres coupled to J-aggregates, which shows the potential of dielectric nanoparticles for strong light–matter interaction at the nanoscale.[53] Despite the novel magnetic hybrid states and anisotropic responses, such an organic−inorganic system is subjected to the photoinstability, incompatibility with semiconductor fabrication process, and uncontrollability of J-aggregates.[28] Alternatively, atomic TMDs provide an ideal platform for the realization of resonance coupling at room temperature in an all-semiconductor system.

In this paper, from both the theoretical and experimental aspects, we for the first time report on the resonance coupling in an all-semiconductor system composed of an individual silicon nanosphere and monolayer $WS_2$ (ML-$WS_2$). The calculated results show that, by coating a silicon nanosphere with 1 nm-thick ML-WS$_2$, strong resonance coupling indicated by a distinct mode splitting and anticrossing behavior with a Rabi splitting energy of 77 meV can occur, which can be attributed to the coherent energy transfer between the magnetic dipole resonance and exciton. Compared with the plasmon−exciton coupling in its plasmonic counterpart, resonance coupling in this all-semiconductor heterostructure is much stronger and less sensitive to the spacing between the nanosphere core and ML-WS$_2$ shell. Furthermore, by utilizing the single-particle dark-field scattering spectroscopy, we can experimentally observe room-temperature quenching dips in the scattering spectra when the silicon nanospheres were placed onto the flat WS$_2$ monolayer grown directly on the SiO$_2$/Si substrate. Due to the small contact area between the nanosphere and flat WS$_2$ monolayer, the resonance coupling in such heterostructure involves only a few excitons, and near-field interaction between the magnetic response and exciton is hindered, thus leads to a relatively weak coupling. Finally, the resonance coupling effect can be actively controlled by tailoring the temperature.

**RESULTS AND DISSCUSSION**

The recent work by Li *et al.* reported the direct growth of layered TMDs on spherical plasmonic nanostructures.[17] Such core−shell heterostructure opens up a new avenue to realize strong light−matter interaction in two-dimensional atomic crystals. In our work, we theoretically conceive an all-semiconductor core−shell heterostructure by coating a silicon nanosphere with ML-WS$_2$ (Figure 1a, insert). The diameter of the silicon nanosphere is 149 nm, and the thickness of the ML-WS$_2$ shell is set to 1 nm to satisfy the experimental value of ML-WS$_2$.[54, 55] The scattering/absorption spectra of small spherical nanoparticles and core−multishell nanostructures can be calculated analytically by Mie theory.[53] Figure 1a gives the theoretical scattering spectrum of the pristine silicon nanosphere, absorption spectrum of the hollow ML-WS$_2$ shell, and the scattering spectrum of the core−shell heterostructure, which are excited by plane wave with liner polarization. The scattering efficiency ($Q_{sca}$, scattering cross section divided by the geometrical cross section) spectrum of the pristine silicon nanosphere exhibits two typical resonant modes in the visible regime, which are dominated by magnetic (around 608 nm) and electric (around 498 nm) dipole modes (Figure 1a, green line and Figure S1 in Supporting Information).[42, 43, 47] For the ML-WS$_2$ hollow shell, we take the absorption into consideration since it is a direct indication of the excitonic transition. The absorption efficiency ($Q_{abs}$, absorption cross section divided by the geometrical cross section) spectrum of ML-WS$_2$ shell shows two prominent absorption peaks at 614 nm and 516 nm (Figure 1a, red line), denoted as A- and B-excitons, respectively. These two excitonic transitions originate from the direct gap transition at the K point with a large strength of spin−orbit coupling (ca. 384 meV).[6, 22] The A-exciton exhibits a narrow linewidth ($\gamma_A$) of 13 meV and large $Q_{abs}$ of 57%, while B-exciton shows a relative broad linewidth ($\gamma_B$) of 60 meV and small $Q_{abs}$ of 32%, which indicates the great potential of A-exciton for the strong light−matter interaction. It is notable that, the $Q_{abs}$ of ML-WS$_2$ shell is relatively large than that of the typical flat monolayer. This is because the surface of spherical shell is four times the area of the geometrical cross section. Therefore, the equivalent $Q_{abs}$ of A- and B-exciton in flat WS$_2$ monolayer is about 14% and 8%, respectively, which agree with the values in the previous report.[22]

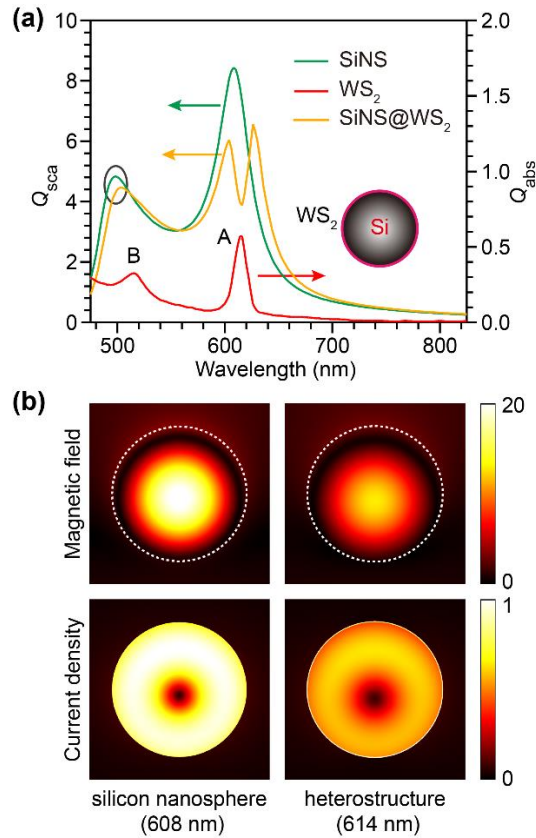

**Figure 1. Resonance coupling in all-semiconductor core−shell heterostructures in free space. (a)** Calculated scattering efficiency spectrum of the pristine silicon nanosphere (green line), absorption efficiency spectrum of the hollow ML-WS$_2$ shell (red line), and the scattering efficiency spectrum of the core−shell heterostructure (orange line). The diameter of the silicon nanosphere is 149 nm, and the thickness of the ML-WS$_2$ shell is set to 1 nm. Insert shows the scheme of silicon nanosphere core−ML-WS$_2$ shell heterostructure. **(b)** Magnetic near-field and current intensity distributions of the pristine silicon nanosphere at magnetic dipole resonance (608 nm) and the core−shell heterostructure at A-excitonic transition (614 nm), respectively. All of the contours are plotted at the central cross section perpendicular to the magnetic field of the incident light.

When the silicon nanosphere is coated with the ML-WS$_2$ shell, due to the relative small oscillator strength and broad linewidth of B-exciton, the interaction of the electric dipole mode with B-exciton is negligible (Figure 1a, orange line). In contrast, a distinct quenching dip emerges in the scattering spectrum of the heterostructure around A-exciton, and two new splitting peaks corresponding to the high- and low-energy modes can form (Figure 1a, orange line), which indicates the resonance coupling between the magnetic dipole mode of silicon nanosphere and A-exciton of the ML-WS$_2$. It is notable that, although there exists an energy difference of 20 meV between the magnetic dipole mode of the pristine silicon nanosphere and A-exciton of the ML-WS$_2$ shell, the interaction is still in resonance, since the magnetic dipole mode redshifts in the heterostructure due to large refractive index of ML-WS$_2$.[56] To further understand the underlying physics of the resonance coupling, we decompose the absorption spectra of the hollow ML-WS$_2$ shell and corresponding heterostructure, and find that the excitonic absorption of A-exciton for hollow ML-WS$_2$ shell is mainly attributed to electric dipole resonant absorption (Figure S2a), while the absorption of the heterostructure at the resonance coupling is almost completely dominated by magnetic dipole resonant absorption (Figure S2b). In addition, we use finite element method (FEM) to supplement the detailed spectral features and the near-field properties of individual nanostructures and corresponding heterostructure. The absorption of the ML-WS$_2$ shell is strongly enhanced in the heterostructure, while that of the silicon nanosphere core is reduced with a mode splitting in the absorption spectrum (Figure S2c). Figure 1b gives the magnetic field and current density distributions of the silicon nanosphere before and after the coating of ML-WS$_2$. Considering the redshift of the magnetic dipole mode of the silicon nanosphere in the heterostructure, the near-field counters of the pristine silicon nanosphere and heterostructure are plotted at 608 nm and 614 nm, respectively. It is clear that, when the heterostructure forms, the magnetic field and current intensity inside the silicon nanosphere is strongly reduced, while the current intensity in the ML-WS$_2$ shell is dramatically enhanced compared with that in hollow ML-WS$_2$ shell in free space (Figure S3 in Supporting Information). The above spectral and near-field properties indicate that, resonance coupling in this all-semiconductor heterostructure is originated from magnetic resonance mediated coherent energy transfer between the magnetic dipole mode and A-exciton.

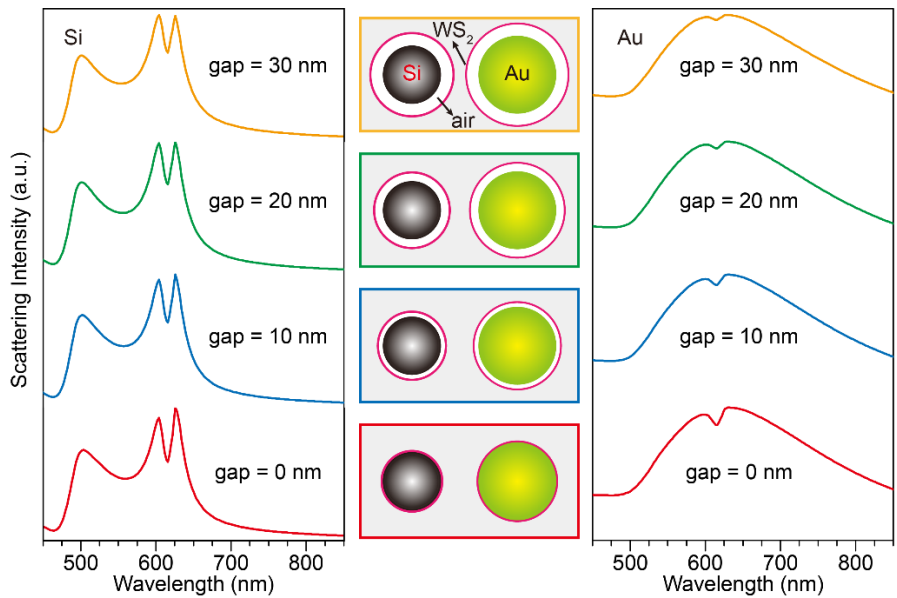

**Figure 2. Calculated scattering spectra of silicon (left) and gold (right) nanospheres coupled to ML-WS$_2$ shell with different separations between the nanosphere core and ML-WS$_2$ shell. The diameter of silicon and gold nanospheres are 149 and 200 nm, respectively. The separation between the core and shell varies from 0 nm to 30 nm. Scheme of the heterostructures are shown in the middle of the panel. The gaps between the core and shell are inserted with air.**

The resonance coupling of an optical cavity with quantum emitters is usually influenced by the separation between these two constituents. When the separation instance is increased, the resonant interactions of excitons with both the plasmonic and dielectric nanostructures will become weaker.[38, 53] However, the dipolar plasmon mode (electric dipole mode) of plasmonic nanostructure and magnetic dipole mode of dielectric nanoparticle are quite different. As is well known, for plasmonic nanostructures, the electric dipole mode is surface type with enhanced electric field distributing at two apexes the plasmonic nanostructures along the incidence polarization, which is sensitive to the surrounding environment. In contrast, for dielectric nanoparticles, the magnetic dipole mode is cavity type where magnetic field is strongly concentrated inside the dielectric nanoparticle, and therefore it is less sensitive to the surrounding environment. In this scenario, the influence of separation on the resonance coupling for plexcitonic and all-semiconductor heterostructures may be different. Figure 2 gives the scattering spectra of the silicon (left) and gold (right) core−ML-WS$_2$ shell heterostructures with different spacings between the core and shell. The diameters of silicon and gold nanospheres are 149 nm and 200 nm, respectively. When the ML-WS$_2$ are coated on the silicon and gold nanospheres, both of these two core−shell heterostructures exhibit pronounced mode splitting in the scattering spectra (red lines). Resonance coupling in silicon nanosphere−ML-WS$_2$ core−shell heterostructure is much stronger than that in the plexcitonic one. The relatively weak interaction of gold nanosphere with ML-WS$_2$ can be attributed to the large optical damping in the gold nanosphere. As the separation between the core and shell is increased, for the gold nanosphere−ML-WS$_2$ heterostructures, mode splitting in the scattering spectra induced by the resonance coupling diminishes quickly and becomes negligible when the spacing is large than 30 nm. On the other hand, for all-semiconductor counterparts, redshift of the magnetic dipole resonance of silicon nanosphere disappears at a large separation, where the influence of the large refractive index of the ML-WS$_2$ can be ignored. Most importantly, resonance coupling in such all-semiconductor heterostructure is less sensitive to the separation between the core and shell, which remains nearly unchanged.

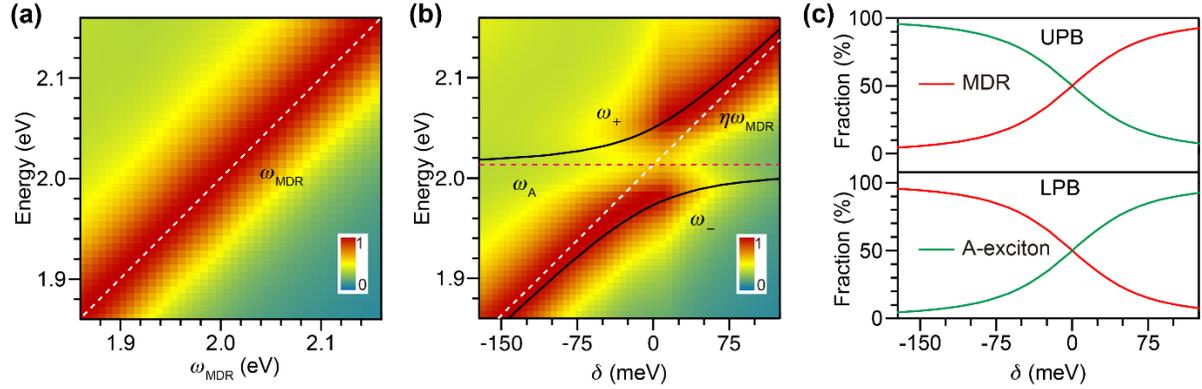

**Figure 3. Anticrossing behavior of the resonance coupling. (a) Normalized scattering energy diagram of silicon nanospheres with diameters varying from 138 nm to 167 nm (color map). White dashed white line presents the magnetic dipole resonance of the pristine silicon nanospheres. Insert shows the color bar. (b) Normalized scattering energy diagram of the heterostructures (color map) corresponding to silicon nanospheres in (a). When the heterostructure forms, the magnetic dipole mode of silicon nanosphere redshifts with the energy modified at a coefficient $\eta$, and the detuning between the effective magnetic dipole mode and A-exciton is $\delta = \eta\omega_{MDR} - \omega_A$. The high-energy upper and low-energy lower polaritons are fitted using coupled harmonic oscillator model (black solid lines) with a distinct anticrossing behavior and a Rabi splitting energy of 77 meV. A-exciton is indicated with red dashed line. Insert shows the color bar. (c) Fraction of A-exciton and MDR for the UPB (upper) and LPB (lower).**

The spectral diagram of the heterostructures with energy detunings can provide further insight into the resonance coupling. By changing the size of silicon nanospheres, resonance energy of the magnetic dipole mode can be facially tuned.[42, 43] Figure 1c shows the normalized scattering spectra of pristine silicon nanospheres with the diameter ranging from 138 nm to 167 nm. The corresponding energy of magnetic dipole resonance ($\omega_{MDR}$) varies from 1.86 eV to 2.16 eV (Figure 3a, white dashed line), which can cover across A-exciton of ML-WS$_2$ shell. Then we calculate the normalized scattering spectra of the corresponding silicon nanosphere−ML-WS$_2$ core−shell heterostructures. It is clear that anticrossing behavior with two prominent branches can be observed on the scattering diagram (Figure 3b, color map), identified as the high-energy upper polariton branch (UPB) and low-energy lower polariton branch (UPB). The dispersion of these two branches can be fitted using a coupled harmonic oscillator model (Figure 2b, black solid lines),

$$\omega_{\pm} = \frac{\omega_{MD} + \omega_A}{2} \pm \frac{\sqrt{\Omega^2 + (\eta\omega_{MDR} - \omega_A)^2}}{2} \tag{1}$$

where $\omega_{MDR}$ and $\omega_A$ are the energy of the uncoupled magnetic dipole resonance and A-exciton, respectively. Parameter $\Omega$ is the Rabi splitting, and $\omega_{\pm}$ are the energies of the hybrid states. Due to the redshift of the magnetic dipole mode of silicon nanosphere (Figure 2b, white dashed line), a modification $\eta$ of resonance energy $\omega_{MDR}$ is applied in the function.[57] Thus, detuning between the coupled magnetic dipole resonance and A-exciton is defined as $\delta = \eta\omega_{MDR} - \omega_A$. The parameter $\eta$ is fitted to be 0.99, which is in accordance with the energy difference of 20 meV between the uncoupled magnetic dipole resonance of silicon nanosphere and excitonic transition of A-exciton shown in Figure 1a. The extracted Rabi splitting energy, $\Omega$, of 77 meV can fulfill the criterion where the strong resonance coupling regime can be reached ($\Omega > \frac{\gamma_{MDR} + \gamma_A}{2}$), since the linewidths of A-exciton, $\gamma_A$, and magnetic dipole resonance of silicon nanosphere, $\gamma_{MDR}$, are 13 meV and 120 meV, respectively. In this regime, strong resonance coupling leads to the formation of hybrid modes with mixed photon-exciton states. The contributions from MDR and A-exciton components for both UPB and LPB are then calculated using the coupled harmonic oscillator model. As shown in Figure 3c, by detuning the MDR to low-energy side of the exciton transition, the LPB is more MDR-like while the UPB more A-exciton-like, and vice versa for detuning to the high energy side. It should be emphasized that although this Rabi splitting energy is smaller than that in the silicon nanosphere−J-aggregates core−shell heterostructure, the criterion for strong resonance coupling is more strict in this all-semiconductor heterostructure, because the linewidths of the A-exciton of ML-$WS_2$ and magnetic dipole mode of silicon nanosphere in free space are much narrower than those of the J-aggregates (50 meV) and magnetic dipole mode of silicon nanosphere in water (200 meV).[53]

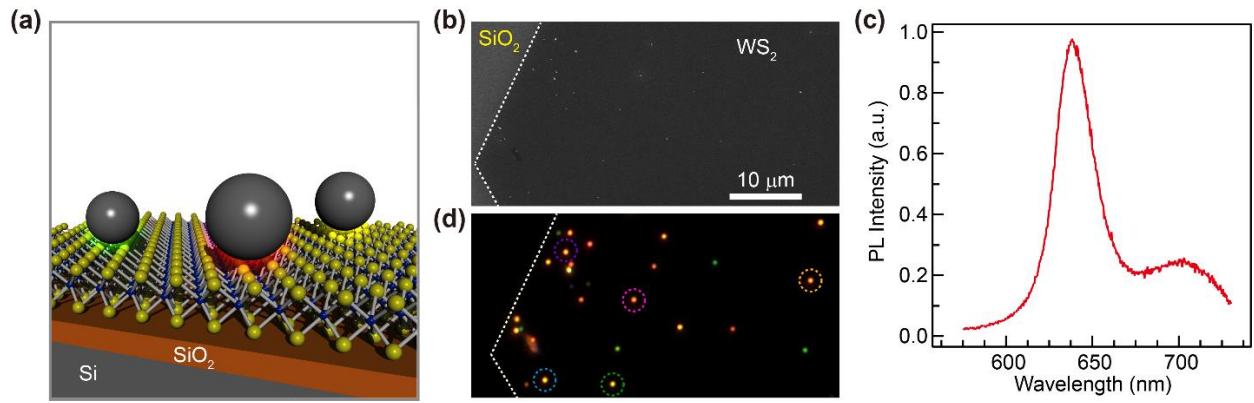

**Figure 4. Silicon nanosphere−flat ML-$WS_2$ all-semiconductor heterostructure. (a) Schematic illustration of the all-semiconductor heterostructure with the individual silicon nanospheres placed onto the surface of a flat $WS_2$ monolayer. (b) SEM image of the $WS_2$ monolayer as grown on $SiO_2$/Si substrate with various silicon nanospheres. The scale bar is 10 μm. (c) Photoluminescence spectrum of the flat $WS_2$ monolayer. (d) Dark-field scattering images of the various heterostructures corresponding to the region in (b). Selected heterostructures are marked with colour dashed circles.**

To experimentally investigate the resonance coupling in all-semiconductor system, an alternative geometry of heterostructure was utilized due to the great challenge for us to fabricate the silicon nanosphere core−ML-WS$_2$ shell heterostructure as proposed in our calculations. As schematically shown in Figure 4a, the silicon nanospheres were placed onto a flat ML-WS$_2$. Therefore, resonance coupling occurs in such heterostructure with a point contact between the silicon nanosphere and flat ML-WS$_2$, which can give rise to different phenomena. In our experiments, large-area WS$_2$ monolayers were grown directly on silicon substrate capped with 200-nm oxide layer by chemical vapor deposition (CVD) methods. The typical thickness of the CVD-WS$_2$ monolayer is ~ 1 nm. Silicon nanospheres with various sizes synthetized by laser ablation in liquid (LAL) were then dispersedly deposited onto as-grown ML-WS$_2$ flake by drop-casting (see SEM image in Figure 4b). In the measurements, characteristic of the excitonic transition in the ML-WS$_2$ flake was conducted by measuring the emission spectrum of the ML-WS$_2$ at room-temperature (Figure 4c). A pronounced emission peak centering at 639 nm (1.94 eV) with a linewidth ($\gamma_A$) of 80 meV dominates the spectrum, which can be attributed to the direct gap transition of the A-exciton.[6, 22] The line shape of this CVD-WS$_2$ is relatively broad with a redshift of 80 meV compared to that of the ML-WS$_2$ obtained by mechanical exfoliation in our calculation, which is consistent with previous reported values.[6, 54] For the study of the resonance coupling in these all-semiconductor heterostructures, single-particle dark-field microscope was utilized to rule out the average effect from the ensemble measurements. The pattern-matching method was employed to correlate the morphology of each heterostructure and its scattering spectrum. Figure 4d shows the dark-field images of various silicon nanospheres on the ML-WS$_2$ marked by colorful dotted circles. In this image, the silicon nanospheres exhibit strong light scattering with various colors due to the strong electric and magnetic dipole resonance.

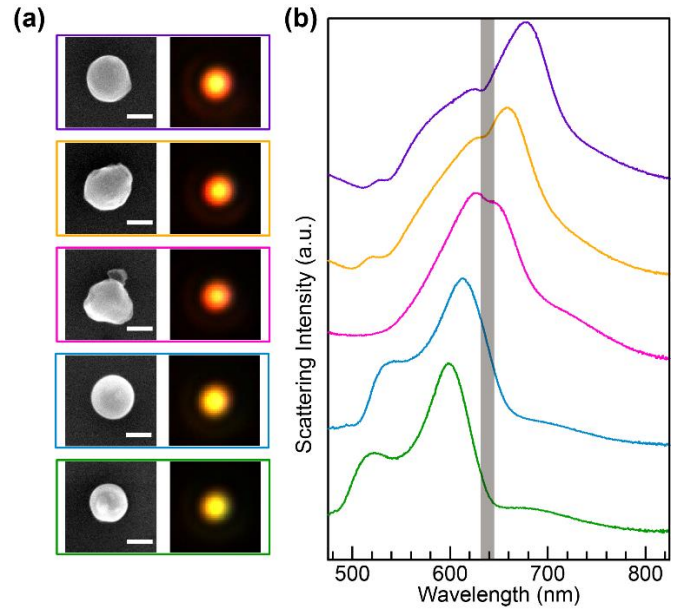

**Figure 5. (a) SEM and dark-field images of the typical heterostructures selected in Figure 4. The diameters of the silicon nanospheres are 153 nm (green), 165 (blue), 169 (pink), 175 nm (orange), and 183 nm (purple), respectively. The scale bar is 100 nm. (b) Dark-field scattering spectra of the silicon nanosphere−flat ML-WS$_2$ heterostructures. The grey vertical line located at 639 nm indicates the A-exciton of the ML-WS$_2$.**

We have measured various silicon nanosphere−ML-WS$_2$ heterostructures with the diameters of the silicon nanospheres ranging from 153 nm to 183 nm. Figure 5a shows the SEM images of five typical silicon nanosphere−ML-WS$_2$ heterostructures and their dark-field images selected in Figure 4d. The scattering spectra of the heterostructures are strongly dependent on the size of the silicon nanospheres (Figure 5b). For the large silicon nanospheres whose magnetic dipole modes are detuned to the low-energy side of the A-exciton of flat ML-WS$_2$ (gray vertical line), we can clearly observe the mode splitting with two pronounced splitting peaks around the magnetic dipole mode of the silicon nanosphere separated by a quenching dip in the dark-field scattering spectrum of each heterostructure (Figure 5, purple, orange, and pink). These two hybrid modes correspond to the high-energy UPB and low-energy LPB, respectively, which are originated from the resonance coupling between the silicon nanospheres and flat ML-WS$_2$. The LPB dominates for the large heterostructure, and blue shifts when size of the silicon nanosphere decreases. Specially, for the silicon nanosphere with a diameter of 169 nm, the magnetic dipole resonance is in resonance with A-exciton, giving rise to almost equivalent magnitudes of LPB and UPB (Figure 5, pink). The electric dipole resonance of this silicon nanospheres is too broad to be distinguished with the magnetic dipole resonance, which can be ascribed to the imperfect shape of the nanosphere. As the sizes of the silicon nanospheres become smaller, the LPB vanishes and only the UPB remains around the A-exciton (Figure 5, blue and green). The two resonance peaks in each of the heterostructure resemble the magnetic and electric dipole modes of the pristine silicon nanospheres.

The above experimental results indicate that, resonance coupling occurs in such silicon nanosphere−flat ML-WS$_2$ heterostructure only when the magnetic dipole mode is tailored to low-energy side of A-exciton. In addition, quenching dips in the dark-field scattering spectra are shallow compared with the theoretical results in the core−shell heterostructures. This is because of the relatively weak resonance coupling in the silicon nanosphere−flat ML-WS$_2$ heterostructure, which can be ascribed to the following reasons. First, the point contact between the silicon nanosphere and flat ML-WS$_2$ gives rise to a rather small contact area, where only a few excitons contribute to the resonance coupling. Second, the magnetic resonance mediated near-field coupling of the silicon nanosphere with ML-WS$_2$ will be hindered in the experimental heterostructure, where strong circular displacement currents can be hardly induced in the flat

ML-WS$_2$. In addition, the linewidth of A-exciton transition of the CVD-WS$_2$ in the experiments (80 meV) is much larger than that in the calculations (13 meV), which is even larger than the energy of Rabi splitting (77 meV) in the theoretical core−shell heterostructures. Therefore, the energy transferred from the silicon nanosphere to the WS$_2$ monolayer will dissipate rapidly, whereby the coupling strength between the magnetic dipole resonance and A-exciton cannot compete with such dissipation.

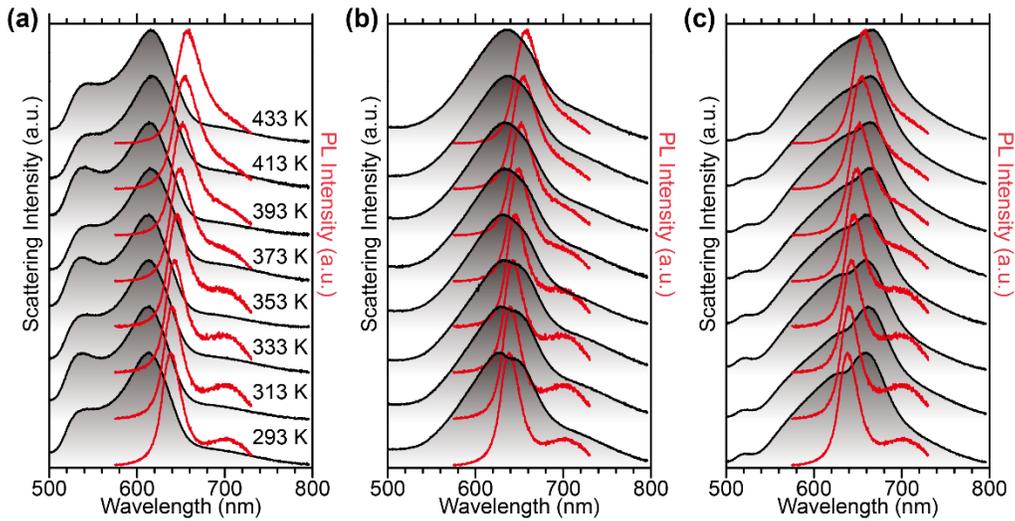

**Figure 6. Temperature-dependent PL spectra of the CVD-WS$_2$ monolayer (red) and scattering spectra from three typical heterostructures (black). The temperature is scanned from 293 K to 433 K.**

In order to further understand the resonance coupling in the all-semiconductor heterostructure, we finally studied the temperature-dependent spectral evolution for both the ML-WS$_2$ and three typical heterostructures. For ML-WS$_2$, the PL spectra are strongly dependent on the external temperature (Figure 6). As the temperature is varied from 293 K to 433 K, the transition energy of A-exciton can be actively tailored from 1.94 eV (639 nm) to 1.88 eV (660 nm), which can be tuned back reversibly (Figure S4). For the smallest silicon nanosphere, the A-exciton of the ML-WS$_2$ is located at the low-energy side of its magnetic dipole mode. When the temperature increases, the A-exciton is detuned to the lower-energy side of the magnetic dipole mode of the silicon nanosphere, therefore the scattering spectra of the heterostructure remain unchanged (Figure 6a). For a larger silicon nanosphere whose magnetic dipole mode is in resonance with the A-exciton at room temperature, the quenching dip resulting from the resonance coupling of the its heterostructure redshifts with the A-exciton. Resonance coupling in this heterostructure becomes weaker because the A-exciton is tuned to the lower-energy side of the magnetic dipole mode of the silicon nanosphere, and disappears gradually when the temperature increment to 433 K. Figure 6c shows the temperature-dependent scattering spectra of the heterostructure with a largest silicon nanosphere. It is clear that, mode splittings can be observed throughout the temperature scanning since the magnetic dipole mode is always at the high-energy side of A-exciton. These results indicate that, resonance coupling in the all-semiconductor silicon nanosphere−ML-WS$_2$ is still robust even over room temperature, and can be actively controlled by temperature scanning.

**CONCULSIONS**

In summary, we have demonstrated, for the first time, resonant photon−exciton coupling in all-semiconductor silicon nanosphere−ML-WS$_2$ heterostructures. In particular, we have theoretically observed strong resonance coupling in the silicon nanosphere core−ML-WS$_2$ shell heterostructure, evidenced by anticrossing behavior in the scattering diagram with a relatively large Rabi splitting of the 77 meV. The calculated spectral and near-field properties show that the resonance coupling results from magnetic response mediated coherent energy transfer between the magnetic dipole resonance of silicon nanosphere core and A-exciton of ML-WS$_2$ shell. Furthermore, resonance coupling in this all-semiconductor core−shell heterostructure is much stronger than that in its plexcitonic counterpart, and less sensitive to the separation between the core and shell. By placing various silicon nanospheres onto a flat ML-WS$_2$ flake, we can also experimentally observe the resonance coupling in the silicon nanosphere−flat ML-WS$_2$ heterostructure manifested by the quenching dips in dark-field scattering spectra at room temperature, although the linewidth of the CVD-WS$_2$ is relatively large and the point contact between the silicon nanosphere and flat ML-WS$_2$ involves a few excitons. Additionally, the resonance coupling is robust over room temperature and can be actively tuned on a single-particle level by temperature scanning. Our results open up a new avenue for exploiting strong light−matter interaction at the nanoscale, and paves the way for room-temperature on-chip polaritonic devices based on all-semiconductor heterostructures.

**METHODS**

**Numerical Simulations.** Numerical simulations were performed using the commercial software package COMSOL Multiphysics v4.3b in the frequency domain. The dielectric function of the ML-WS$_2$ was adopted from previous reported values.[56] The dielectric functions of silicon and gold were employed according to previous measurements.[58] A linearly polarized plane wave with wavelength ranging from 475 to 825 nm was launched into a box containing the target nanostructure. A maximum mesh size of 5 nm was applied to the nanostructure, and the surrounding medium were divided using fined meshes. The surrounding medium was set as vacuum with refractive index of 1.0. Perfectly matched layers were used at the boundary to absorb the scattered radiation in all directions. The absorption spectra of the core and shell were obtained by integrating the power loss density over their volumes, respectively. The magnetic field and current density distributions were inspected at the central cross section perpendicular to the magnetic field of the incident light.

**Sample preparation.** In our study, the monolayer WS$_2$ was grown directly onto the silicon substrate capped with 300-nm oxide layer according to procedures reported previously.[59] The silicon nanospheres with various diameters were synthesized using the femtosecond laser ablation in liquid. Single-crystalline silicon wafer was used as target and immersed in deionized water. Legend Elite Series ultrafast laser (Coherent Inc.) was utilized as the ablation source in our experiment (wavelength 800 nm). The pulse width is 35 fs, with the energy of a single pulse of 4 mJ and the repetition rate of 1 kHz. The silicon nanosphere colloidal solution can then be obtained after laser ablating of the silicon wafer in water, and are thereafter drop-casted onto the monolayer WS$_2$ flake. Various heterostructures can be obtained after the deposit was dried naturally under ambient conditions.

**Dark-Field Scattering Imaging and Spectroscopy.** The scattering spectra of the heterostructures were recorded on a dark-field optical microscope (Olympus BX51) that was integrated with a quartz-tungsten-halogen lamp (100 W), a monochromator (Acton SpectraPro 2360), and a charge-coupled device camera (Princeton Instruments Pixis 400BR_eXcelon). The camera was thermoelectrically cooled down to −70 ºC during the measurements. A dark-field objective (100×, numerical aperture 0.80) was employed for both illuminating the heterostructures with the white excitation light and collecting the scattered light. The incidence angle is ~ 55°. The scattered spectra from the heterostructures were corrected by first subtracting the background spectra taken from the adjacent regions without nanostructures and then dividing them with the calibrated response curve of the entire optical system. Color scattering images were captured using the color digital camera (ARTCAM-300MI-C, ACH Technology Co., Ltd., Shanghai) mounted on the imaging plane of the microscope.

**Characterizations.** The photoluminescence spectra of the ML-WS$_2$ were acquired using a Renishaw inVia Reflex system with a dark-field microscope (Leica). The excitation laser of 532 nm was focused onto the samples with a diameter of ~ 1 μm through a 50× objective (numerical aperture 0.80). SEM imaging was performed using an FEI Quanta 450 microscope.

AUTHOR INFORMATION

**Corresponding Author**

*E-mail: stsxns@mail.sysu.edu.cn; stsdsz@mail.sysu.edu.cn; chenhj8@mail.sysu.edu.cn.

**Author Contributions**

The manuscript was written through contributions of all authors. All authors have given approval to the final version of the manuscript.

**Notes**

The authors declare no competing financial interest.

ACKNOWLEDGMENT

This work was financially supported by the National Natural Science Foundation of China (Grant Nos. 51290271, 11474364), the National Key Basic Research Program of China (Grant Nos. 2013CB933601, 2013YQ12034506), the Guangdong Natural Science Funds for Distinguished Young Scholars (Grant No. 2014A030306017), Pearl River S&T Nova Program of Guangzhou (Grant Nos. 201610010084), and the Guangdong Special Support Program.